\documentclass[twocolumn]{aastex62}

\usepackage{amsfonts}
\usepackage{placeins}
\usepackage{amsmath}
\usepackage{nicefrac}

\mathchardef\mhyphen="2D
\newcommand{\Av}{\ensuremath{A_{\mathrm V}}}
\newcommand{\Avo}{\ensuremath{A_\textrm{Vmax}}}
\newcommand{\sigmamax}{\ensuremath{\sigma_\mathrm{max}}}

\newcommand{\LIR}{\ensuremath{L_\mathrm{IR}}}
\newcommand{\LUV}{\ensuremath{L_\mathrm{UV}}}

\shorttitle{A Geometric Model for Dust Attenuation in Star-Forming Galaxies}
\shortauthors{Zuckerman et al.}

\begin{document}

\title{Reproducing the $UVJ$ Color Distribution of Star-forming Galaxies at $0.5 < z < 2.5$ \\with a Geometric Model of Dust Attenuation}

\correspondingauthor{Leah D.\ Zuckerman}
\email{Leah\_Zuckerman@Brown.edu}

\author[0000-0002-6824-8198]{Leah D.\ Zuckerman}
\affiliation{Center for Astrophysics $|$ Harvard \& Smithsonian, 60 Garden St., Cambridge, MA 02138, USA}
\affiliation{Brown University, Providence, RI 02912, USA}

\author{Sirio\ Belli}
\affiliation{Center for Astrophysics $|$ Harvard \& Smithsonian, 60 Garden St., Cambridge, MA 02138, USA}

\author[0000-0001-6755-1315]{Joel Leja}
\affil{Department of Astronomy \& Astrophysics, The Pennsylvania State University, University Park, PA 16802, USA}
\affil{Institute for Computational \& Data Sciences, The Pennsylvania State University, University Park, PA, USA}
\affil{Institute for Gravitation and the Cosmos, The Pennsylvania State University, University Park, PA 16802, USA}

\author{Sandro\ Tacchella}
\affiliation{Center for Astrophysics $|$ Harvard \& Smithsonian, 60 Garden St., Cambridge, MA 02138, USA}
\affiliation{Department of Physics, Ulsan National Institute of Science and Technology (UNIST), Ulsan 44919, Republic of Korea}

\begin{abstract}

We analyze the distribution of rest-frame $U-V$ and $V-J$ colors for star-forming galaxies at $0.5 < z < 2.5$. Using stellar population synthesis, stochastic star formation histories, and a simple prescription for the dust attenuation that accounts for the shape and inclination of galaxies, we construct a model for the distribution of galaxy colors. With only two free parameters, this model is able to reproduce the observed galaxy colors as a function of redshift and stellar mass remarkably well. Our analysis suggests that the wide range of dust attenuation values measured for star-forming galaxies at a given redshift and stellar mass is almost entirely due to the effect of inclination; if all galaxies at a given stellar mass were observed edge-on, they would show very similar dust attenuation. This result has important implications for the interpretation of dust attenuation measurements, the treatment of UV and IR luminosity, and the comparison between numerical simulations and observations.

\end{abstract}

\keywords{High-redshift galaxies, Galaxy evolution, Galaxy colors, Two-color diagrams, Extinction}

\section{Introduction}
\label{sec:intro}

Much of our knowledge about the physical properties of distant galaxies derives from fitting models to multi-band photometric observations \citep[][and references therein]{Conroy_2013}. By combining high-quality panchromatic photometry with modern fitting techniques it is now possible to study individual galaxies in great detail, and constrain realistic physical models with a large number of free parameters \citep[e.g.,][]{Johnson_2021}.

A complementary approach is to consider a small number of filters or colors and study their \emph{distribution} for a galaxy population. This is typically done using color-color diagrams; among these, the most popular one is the $UVJ$ diagram, i.e. rest-frame $U-V$ versus $V-J$ \citep{Wuyts_2007, Williams_2009}. The combination of these two colors is able to break the degeneracy between age and dust reddening, and is therefore particularly effective at separating quiescent and star-forming galaxies. However, the $UVJ$ diagram has been used for more than merely grouping galaxies into two categories. For example, $UVJ$ colors have been used to infer the star formation rate and dust attenuation for star-forming systems \citep{Fang_2018}, and the stellar ages for quiescent systems \citep{Belli_2019}. This is possible because galaxies follow scaling relations, so that their properties are tightly correlated and span only a small region of the available parameter space. In other words, the $UVJ$ diagram reveals more about the physical properties than what can be gathered by the analysis of the $U-V$ and the $V-J$ colors alone \citep{Leja_2019}. 

The regularity observed in the distribution of UVJ galaxy colors, however, is still poorly understood. For example, the most advanced simulations of galaxy formation still struggle to reproduce the observed color distribution of star-forming galaxies \citep[e.g.,][]{Donnari_2019, Akins_2021}. The goal of the present work is to develop a simple, empirical model that is able to reproduce the observed distribution of star-forming galaxies on the $UVJ$ diagram as a function of redshift and stellar mass.

\section{Data} \label{sec:data}

We use the 3D-HST catalog \citep{Brammer_2012, Skelton_2014, Momcheva_2016} of galaxies detected in \emph{Hubble Space Telescope} (HST) imaging from the five fields of the CANDELS survey \citep{Grogin_2011, Koekemoer_2011}. 
We select galaxies with redshift (obtained by a joint fit to grism spectroscopy and photometry) in the range $0.5 < z < 2.5$ and with stellar mass (derived from a fit to the photometry) in the range $9.5 < \log(M_\ast / M_\odot) < 11$. This selection yields a sample of 19553 galaxies that is more than 90\% complete \citep{Tal_2014}. We then remove galaxies with bad photometry (via the \texttt{use\_phot} flag provided in the catalog), bad fit ($\chi^2 > 5$), or a bad measurement of the axis ratio $q$ in the \citet{vanderWel_2014} catalog. This leaves us with a final sample of 15073 galaxies.

For these galaxies we retrieve the rest-frame $U-V$ and $V-J$ colors from the 3D-HST catalog, which are derived from the observed photometry using the EAZY code \citep{Brammer_2008}. We show the distribution of the sample on the $UVJ$ diagram in Figure~\ref{fig:3DHSTbyAxRatio}, split into three stellar mass bins and four redshift bins, and color-coded by axis ratio.
The black line shows the definition of the quiescent box: $U-V > 1.3$ and $U-V > 0.88(V-J)+0.59$ \citep{Muzzin_2013, Belli_2019}. We use these criteria to select the star-forming population, which lies below the black line.

\section{Model} \label{sec:methods}

In order to understand the distribution of galaxies on the $UVJ$ diagram we develop a simple Monte Carlo simulation (described below) which includes stochastic star formation histories (SFHs), stellar population synthesis, and a geometric model for the effect of dust attenuation.

\subsection{Star Formation Histories} \label{subsec:SFHs}

We adopt the stochastic model of \cite{Caplar_2019} to generate SFHs that are realistic but require minimal physical assumptions. 
In this model, galaxies are assumed to fluctuate about the star-forming main sequence, which describes the typical star formation rate as a function of stellar mass \citep{Noeske_2007}. Since we are interested only in galaxy colors and not in the absolute fluxes, the normalization of the main sequence does not matter and we assume that the typical star formation rate for all galaxies, at all redshifts, is $1~\mathrm{M}_{\odot}~\mathrm{yr}^{-1}$. 

This approach neglects the fact that the normalization of the main sequence declines with cosmic time. As we explain below, this assumption should not have a strong effect on the distribution of synthetic colors. Other work supports this model, for example Iyer et al. (2020) find that at fixed stellar mass, galaxies in the IllustrisTNG simulation have roughly a constant star formation rate (except at high stellar masses where the effect of quenching, which we do not consider in this work, is important).

In the \cite{Caplar_2019} model, a stochastic SFH is fully characterized by the power spectrum density (i.e., the power as a function of frequency) of its fluctuations around the main sequence. The power spectrum density is assumed to be constant on timescales longer than a break timescale $\tau_{\rm x}$, below which the power spectrum follows a power law with exponent $\alpha=-2$, corresponding to a damped random walk. \citet{Tacchella_2020} show that this power spectrum density can be obtained by solving the gas regulator model \citep{Lilly_2013} when the gas accretion is a white noise process. Moreover, a broken power law with a break timescale of approximately 1 Gyr is consistent with the results of numerical simulations \citep{Iyer_2020}. For this reason we adopt a fixed value of $\tau_x$ = 1 Gyr.

We begin the Monte Carlo simulation by generating 100 stochastic SFHs with a time sampling of 1 Myr and a duration equal to the current age of the universe. This library of SFHs constitutes the core of our model.

\begin{figure*}[ht]
\centering
\includegraphics[width=\textwidth]{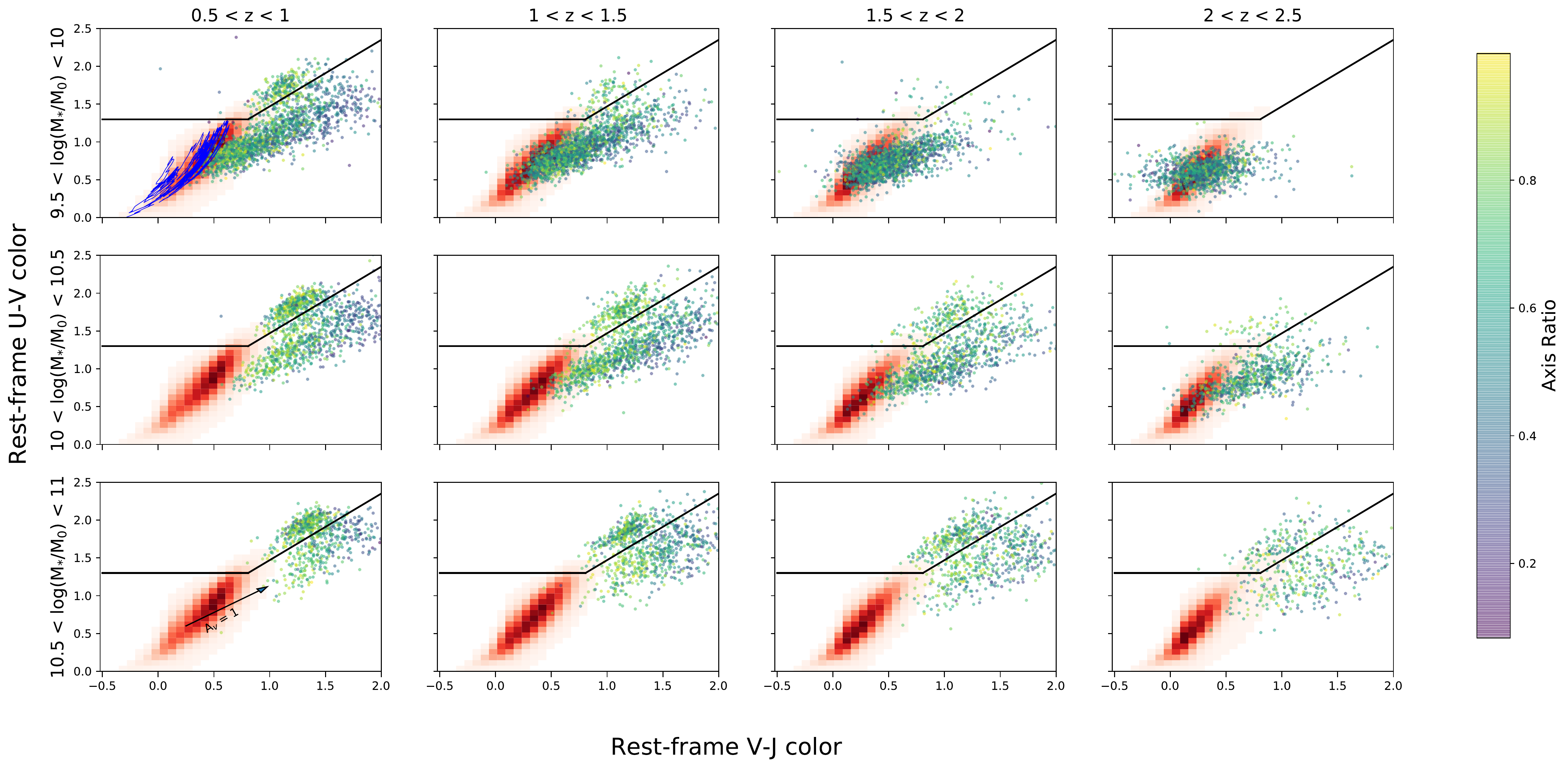}
\caption{Rest-frame $U-V$ versus $V-J$ colors for galaxies in the CANDELS fields, split into mass and redshift bins. Galaxies are color-coded according to their axis ratio. The blue line in the first panel is a single dust-free synthetic galaxy track; the red distribution is the histogram of many individual tracks like this one. The black arrow in the bottom left panel represents the color shift due to an attenuation $A_V$ of one magnitude, assuming the \citet{Calzetti_2000} law.}
\label{fig:3DHSTbyAxRatio}
\end{figure*}

\subsection{Stellar Population Synthesis} \label{subsec:FSPS}

To produce synthetic galaxy colors we rely on the Flexible Stellar Population Synthesis (FSPS) library \citep{Conroy_2009, Conroy_2010}. We assume a \citet{Kroupa_2001} initial mass function, zero dust attenuation, and we include nebular emission as described in \cite{Byler_2017}. Then we calculate the synthetic photometry in the rest-frame Johnson $U$, Johnson $V$, and 2MASS $J$ filters (the same ones used for the observed galaxies) at each time step of our stochastic SFHs. The blue line in the first panel of Figure~\ref{fig:3DHSTbyAxRatio} shows an example of synthetic evolutionary track obtained in this way. The track features a high degree of variability, reflecting the fact that the SFH is, by construction, a stochastic process. The model galaxy reaches the bluest colors when experiencing a peak in star formation rate, while in times of relative quiescence its colors become redder.

Observations of galaxies, however, are unable to constrain the track of individual systems across cosmic time, but can only measure the color distribution for a galaxy population. We therefore build a synthetic population of galaxies by drawing 10 different random values of stellar metallicity (which has a smaller effect on the colors compared to the SFH), uniformly distributed over $-0.5 < \log Z/Z_\odot < 0.2$, for each of the 100 stochastic SFHs. This gives a sample of 1000 synthetic galaxy tracks, each of which includes hundreds of points on the $UVJ$ diagram. The resulting two-dimensional distribution is shown in red in Figure~\ref{fig:3DHSTbyAxRatio}. In each bin we only consider ages between zero and the age of the universe at that redshift.
The model distributions are quite similar at all redshifts despite the substantial difference in the maximum stellar age considered. This is not surprising, since the photometry of star-forming galaxies is mainly driven by the younger, more luminous stars, and is not very sensitive to the past history. It is likely, then, that the simplifying assumption of a non-evolving main sequence has a negligible effect on our conclusions.

\begin{figure*}[ht]
\centering
\includegraphics[width=\textwidth]{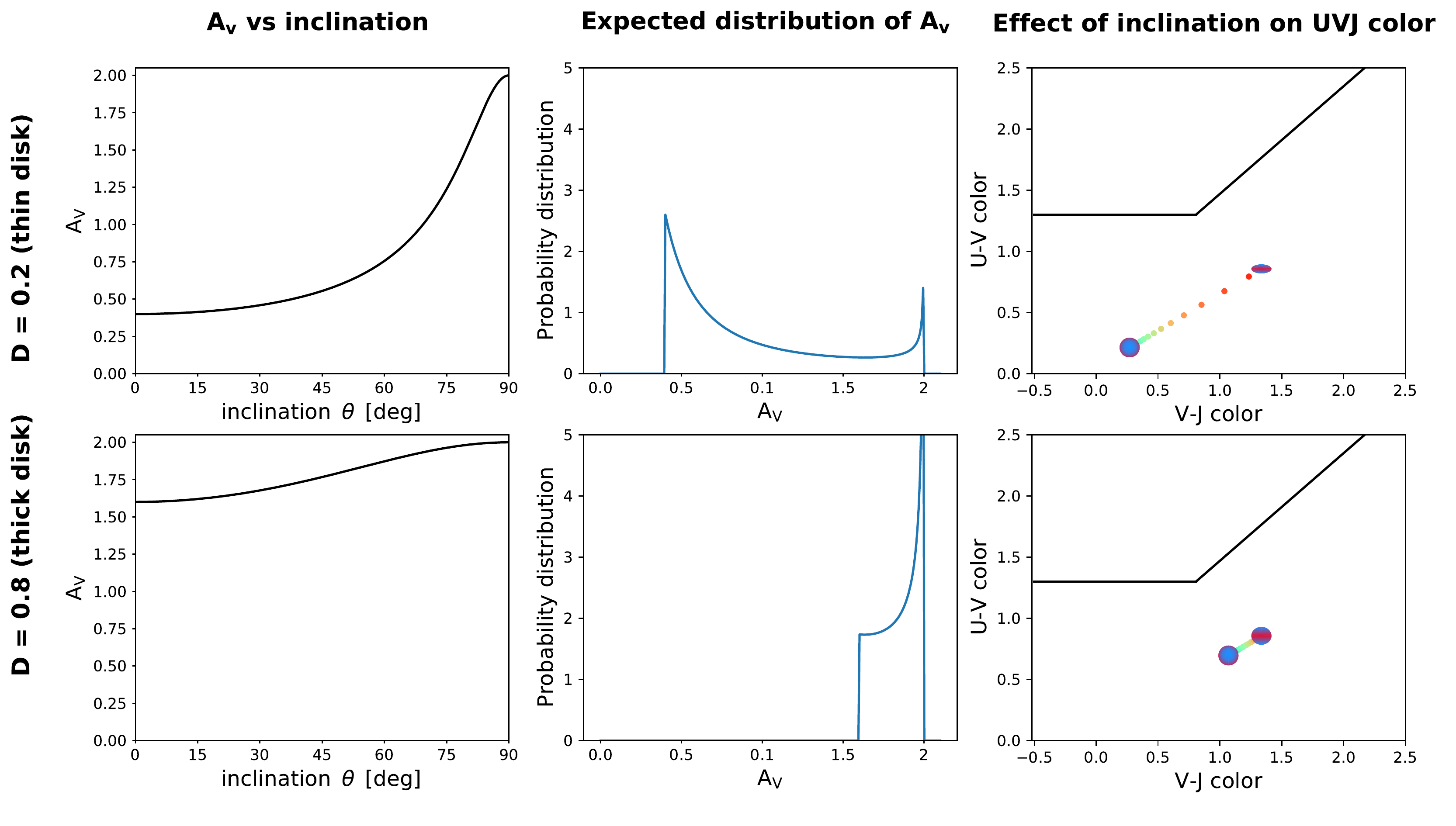}
\caption{Illustration of our geometric model of dust attenuation for two examples: a thin disk with $D=0.2$ (top row) and a thick disk with $D=0.8$ (bottom row); in both cases $\Avo = 2$. \emph{Left:} dust attenuation as a function of the inclination angle. \emph{Center:} distribution of dust attenuation for an ensemble of galaxies with random inclinations. \emph{Right:} variation of the $UVJ$ colors as a function of inclination angle; where the 2D representations of galaxies have inclination angles of 10$^{\circ}$ and 90$^{\circ}$.}
\label{fig:geometricModel}
\end{figure*}

\subsection{Dust Attenuation}
\label{subsec:Treatment of Dust}

The synthetic distributions in Figure~\ref{fig:3DHSTbyAxRatio} clearly do not provide a good fit to the data, and this is because the effect of dust reddening has so far been neglected. 
We illustrate the expected color shift caused by an attenuation of one magnitude in the $V$ band ($A_V=1$) in the bottom left panel of Figure~\ref{fig:3DHSTbyAxRatio}. The exact direction of this dust arrow depends on the dust attenuation law; we assume the widely used \citet{Calzetti_2000} law. This attenuation law appears to produce a color shift along the direction connecting the synthetic colors to the observed ones. For this reason, and to keep the model as simple as possible, we assume the Calzetti law for all the synthetic galaxies, even though in principle some variation between galaxies is expected \citep{Salim_2020}.

Dust attenuation is clearly fundamental in shifting galaxies along the diagonal direction of the $UVJ$ diagram. However, Figure~\ref{fig:3DHSTbyAxRatio} also shows a strong trend of the observed axis ratio along the diagonal direction. In particular, among star-forming galaxies those that are found at the reddest edge of the UVJ distribution are mostly edge-on, a fact first noticed by \citet{Patel_2012}.This suggests that the dust attenuation in each galaxy strongly depends on the geometry of the system, consistent with previous work \citep[e.g.,][]{Giovanelli_1995}.

Several studies have sought to determine the effect of inclination in a variety of ways, for example using the thickness of the Tully-Fisher relation \citep[e.g.,][]{Verheijen_2001}, galaxy SED fitting \citep[e.g.,][]{Kauffmann_2003}, or by  examining how the color of individual galaxies change as a function of inclination \citep[e.g.,][]{Maller_2009}.  Instead, in this work we choose to capture this correlation with a simple geometric model in which each galaxy is represented by an oblate spheroid with half-radii $a$, $b$, $c$ with $a=b>c$, and where $D=c/a$ is the galaxy thickness. We define the inclination angle $\theta$ so that a fully face-on galaxy has $\theta=0^{\circ}$ and a fully edge-on galaxy has $\theta=90^{\circ}$.
We then make the assumption that the dust is uniformly distributed in the galaxy, so that the optical depth as a function of inclination can be calculated from the geometry of the model: $\tau(\theta) \propto a \, (\sin^2 \theta+D^{-2}\cos^2 \theta)^{-1/2}$. By assuming that the attenuation \Av\ is proportional to the optical depth along the line-of-sight (and therefore neglecting the emission from stars that are mixed with the dust), we obtain the following relation:
\begin{equation} 
\Av(\theta) = \frac{\Avo}{\sqrt{\sin^2 \theta+D^{-2}\cos^2 \theta}} \; ,
\label{eq:attenuation}
\end{equation}
where \Avo\ is the attenuation when the galaxy is viewed edge-on.

To illustrate the effect of galaxy geometry on the dust attenuation, in Figure~\ref{fig:geometricModel} we show the results of our model for a thin disk ($D=0.2$, top row) and for a thick disk ($D=0.8$, bottom row), assuming $\Avo = 2$. The first column illustrates the relation between \Av\ and inclination given in Equation~\ref{eq:attenuation}. As expected, the inclination effect is strong for thin galaxies and less important for thicker galaxies; in the limit of a spherical galaxy ($D=1$) there would be no difference between face-on and edge-on attenuation. The central panels show the expected distribution of dust attenuation for an ensemble of identical galaxies observed from random lines of sight. Thin galaxies feature a peak around the minimum value of attenuation (which, in our model, is given by $\Av = D \cdot \Avo$), while thick galaxies experience a substantially smaller dynamic range in attenuation. Finally, the panels on the right show the effect of dust attenuation on the $UVJ$ colors. We conclude that, in this simple model, the galaxy shape (i.e. the thickness parameter $D$) and the dust content (parameterized with \Avo) play a fundamental role in determining the distribution of a galaxy population on the $UVJ$ diagram.

Our geometric model of dust attenuation is substantially more flexible than the often-used slab model (where $\tau \propto 1 / \cos \theta$), while at the same time being conceptually simpler and requiring less assumptions than some of the more advanced models \citep[e.g.,][]{Chevallard_2013}. The key feature of our approach is the presence of a free parameter describing the galaxy thickness; see  \citet{Padilla_2008} for a similar approach.

\begin{figure*}[ht]
\centering
\includegraphics[width=\textwidth]{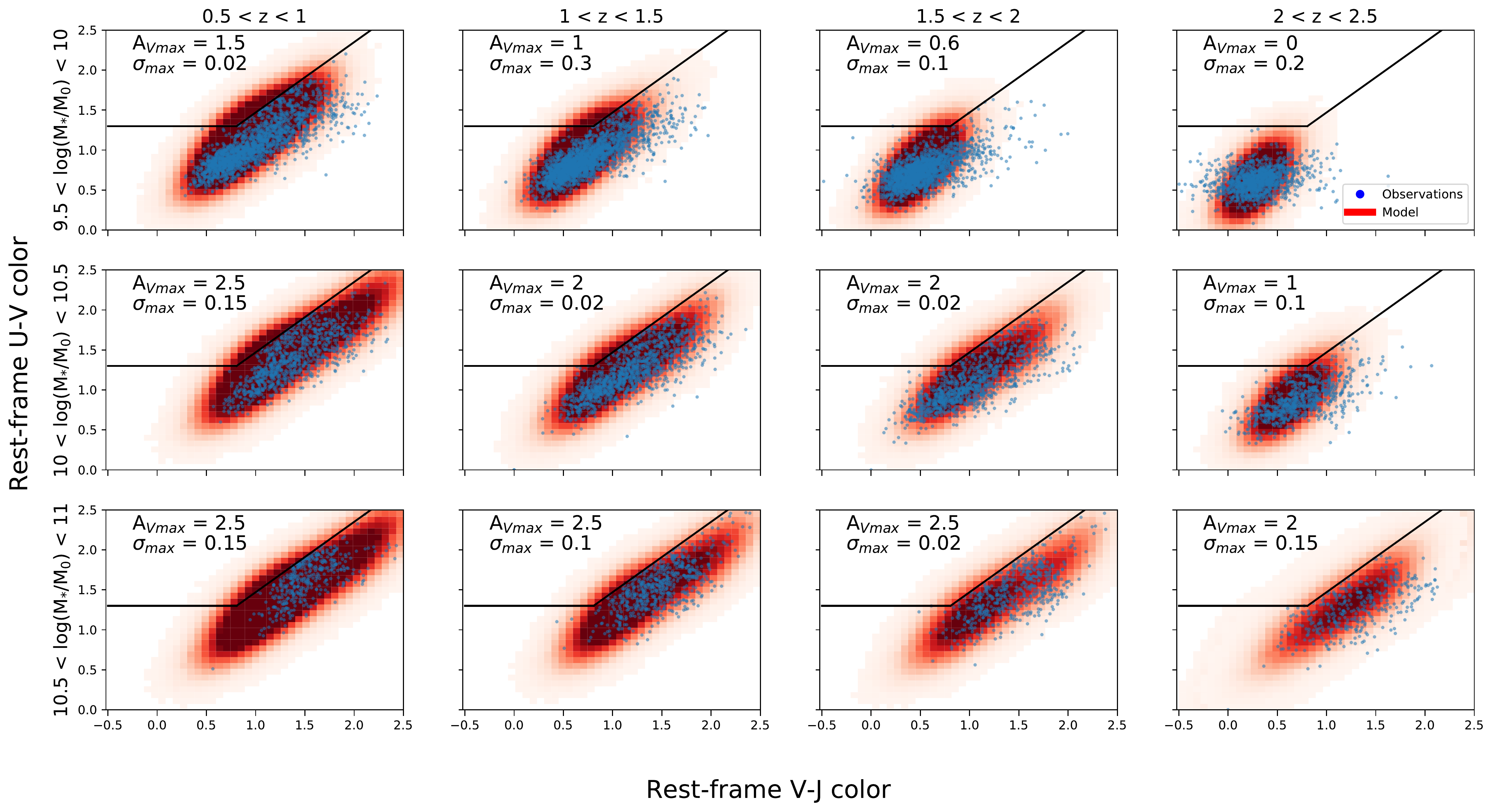}
\caption{Best-fit models (red) and observed galaxies (blue; only star-forming systems are shown) on the $UVJ$ diagram. The best-fit values of the two model parameters, \Avo\ and \sigmamax, are listed in each bin.}
\label{fig:bestfit}
\end{figure*}

\section{Results}
\label{sec:results}

\subsection{Monte Carlo Modeling of Galaxy Populations}

Central to our analysis is the application of the geometric model to the synthetic $UVJ$ colors: this requires the parameters $\theta$, $D$, and \Avo\ for each synthetic track.
Since galaxies are randomly oriented in space, we draw the inclination $\theta$ from a random distribution uniform in $\cos \theta$. 
For the galaxy thickness $D$ we use the results of \citet{vanderWel_2014}, who derived the intrinsic shape of star-forming galaxies from the distribution of axis ratios measured for the 3D-HST sample. They provide the mean and standard deviation for the distribution of $D$, assumed to be Gaussian, in each redshift and stellar mass bin. We use these distributions to draw a random value of $D$ for each synthetic track. Finally, we assume that the \Avo\ parameter of all the galaxies in a given bin follows a Gaussian distribution with free mean and dispersion values, which represent the only two free parameters of our Monte Carlo model.

\subsection{Model Fitting}

For each bin in mass and redshift, we divide the $UVJ$ plane into $n$ cells and calculate the observed probability distribution $p_i$ (where $i = 1, 2, ..., n$ is the cell index) representing the fraction of observed galaxies that lie in each cell. Then we generate a grid of Monte Carlo models by varying the two free parameters: $\Avo =$ 0, 0.3, 0.6, 0.9, 1, 1.5, 2, 2.5 and $\sigmamax =$ 0.02, 0.06, 0.1, 0.15, 0.2, 0.3, 0.4, 0.5, 0.6, controlling respectively the mean and dispersion of the maximum (i.e., edge-on) dust attenuation. At each point of the model grid we calculate the model probability distribution $q_i(\Avo, \sigmamax)$ using the same $UVJ$ cells as for the observations. We also include a small Gaussian scatter of width 0.1 magnitude that represents the typical observational uncertainty on the colors. We then compare the data and the model using the K-L divergence \citep{Kullback_1951}: 
\begin{equation}
D_{KL}(p_i,q_i) = \left\{
        \begin{array}{ll}
            p_i \log(p_i / q_i) & \quad \mathrm{if \ } p_i>0,\ q_i>0 \\
          0 & \quad \mathrm{if \ }  p_i=0,\ q_i>0 \\
            \infty & \quad \mathrm{if \ }  p_i>0,\ q_i=0
        \end{array}
    \right.
\end{equation}

The K-L divergence is infinity for cells where no model galaxies, but at least one observed galaxy is present. In principle this eliminates models that are clearly inconsistent with the data, but we need some flexibility to account for systematic uncertainties (such as photometric redshift errors). We therefore assign, in each model, a small probability to all $UVJ$ cells that represents the rare possibility of substantial outliers. We can then calculate the total K-L divergence between the data and a specific model by integrating over all the $UVJ$ cells:
\begin{equation}
D_{KL} = \sum_{i=1}^{n} D_{KL}(p_{i},q_{i}) \; .
\end{equation}
For a large number of galaxies $N$, $D_{KL} = -1/N \log \mathcal{L}$, where $\mathcal{L}$ is the likelihood of obtaining the data if the model is correct \citep[see, e.g.,][]{Shlens_2014}. Thus, the model that minimizes the K-L divergence is also the model that maximizes the likelihood, and we call this the ``best fit''.

\begin{figure}[ht]
\centering
\includegraphics[width=0.45\textwidth]{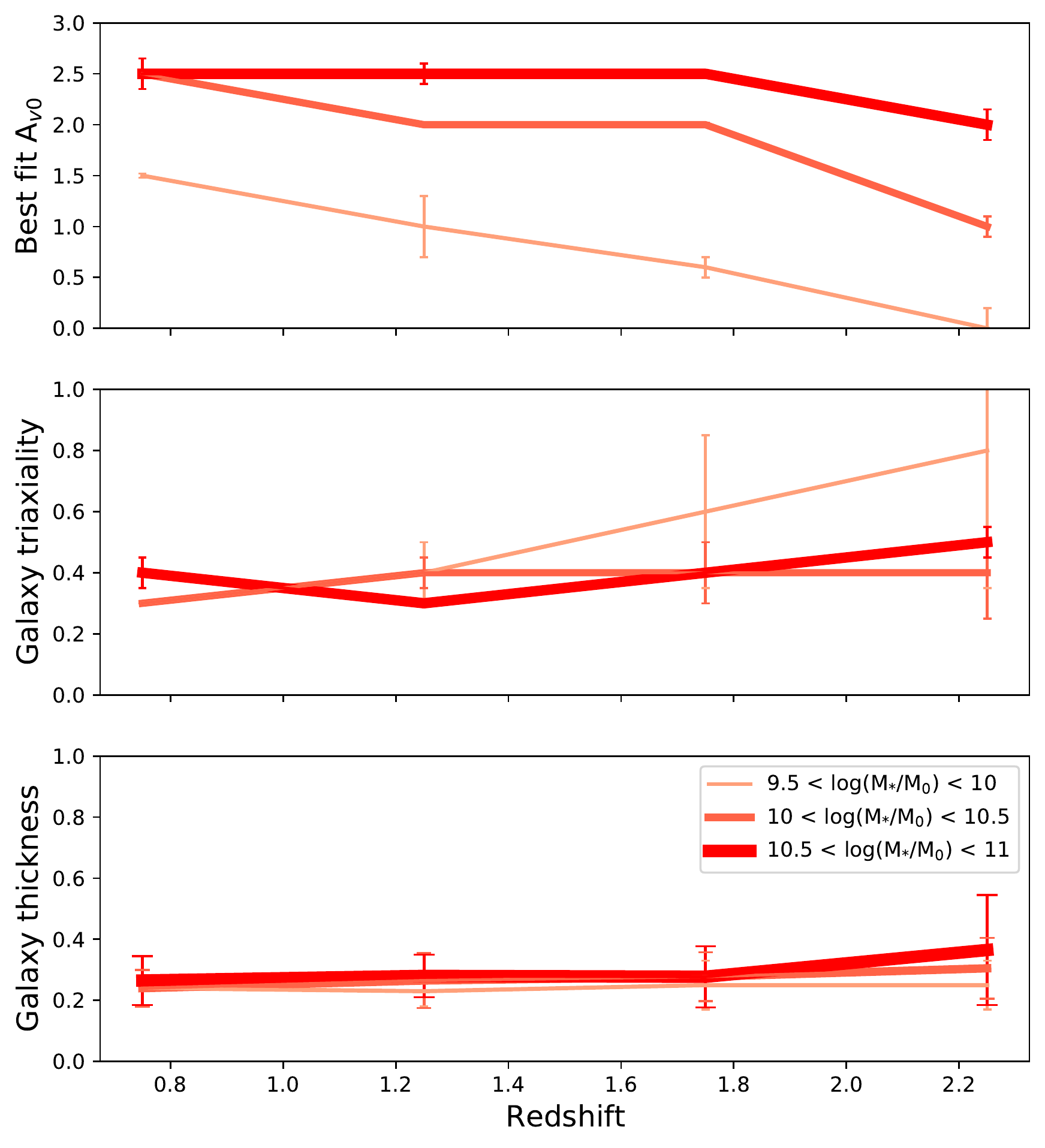}
\caption{The top panel shows the best-fit \Avo\ values as a function of redshift, and the vertical bars represent \sigmamax, the scatter on \Avo\ in each galaxy population. In the second and third panels we show the redshift evolution of the galaxy triaxiality and thickness measured by \cite{vanderWel_2014}, again with vertical bars showing the standard deviation on each parameter.}
\label{fig:paramsOverZ}
\end{figure}

\subsection{Dust Attenuation across Mass and Redshift}

We show the results of the fit in Figure~\ref{fig:bestfit}. In each redshift and stellar mass bin the observed galaxies are shown in blue and the best-fit model in red. Considering that our simple model has only two free parameters, it is able to reproduce the data remarkably well across the mass and redshift range. 
The best-fit values of the model parameters are listed in each panel, and are also plotted in the top panel of Figure~\ref{fig:paramsOverZ} as a function of redshift, with the vertical bars representing the scatter \sigmamax. 
We find that the scatter is low compared to \Avo, particularly for systems with intermediate and high stellar masses. This means that, at fixed mass and redshift, all star-forming galaxies have roughly the same intrinsic amount of dust, and the wide observed range in \Av\ and color is mostly due to inclination, confirming the result of \citet{Patel_2012}. We also find that the dust content of galaxies grows with cosmic time and stellar mass, in agreement with previous studies \citep[e.g.,][]{Fang_2018}.

The bottom two panels of Figure~\ref{fig:paramsOverZ} show the redshift trend for the galaxy triaxiality and thickness measured by \cite{vanderWel_2014}. Most galaxies are relatively thin disks with $D \sim 0.25$ and a small scatter. However, low-mass systems at high redshift are also highly triaxial, which suggests that our oblate spheroid model (which has zero triaxiality) is not appropriate in this regime. 
Interestingly, Figure~\ref{fig:bestfit} shows that the model does not fit this population of low-mass, high-redshift galaxies well, with the best-fit model extending on the diagonal rather than the horizontal direction of the $UVJ$ diagram. However, the fact that the best-fit dust attenuation is negligible for this population may also indicate a problem with the assumed SFH, such as the strength or timescale of its stochastic fluctuations. 

Additionally, our model appears insufficient to reproduce the observations at the opposite end of the parameter space, i.e. for high-mass, low-redshift galaxies. In this case the model produces too many galaxies with low \Av. This is likely connected to the simplistic assumptions regarding the spatial distribution of stars in our model. If one were to consider the contribution from stars that are uniformly mixed with the dust, then the model would need a much higher \Avo\ to reproduce the dustier systems, and as a consequence the face-on galaxies would also appear more attenuated.

\section{Discussion}
\label{sec:discussion}

\begin{figure*}[ht]
\centering
\includegraphics[width=0.4\textwidth]{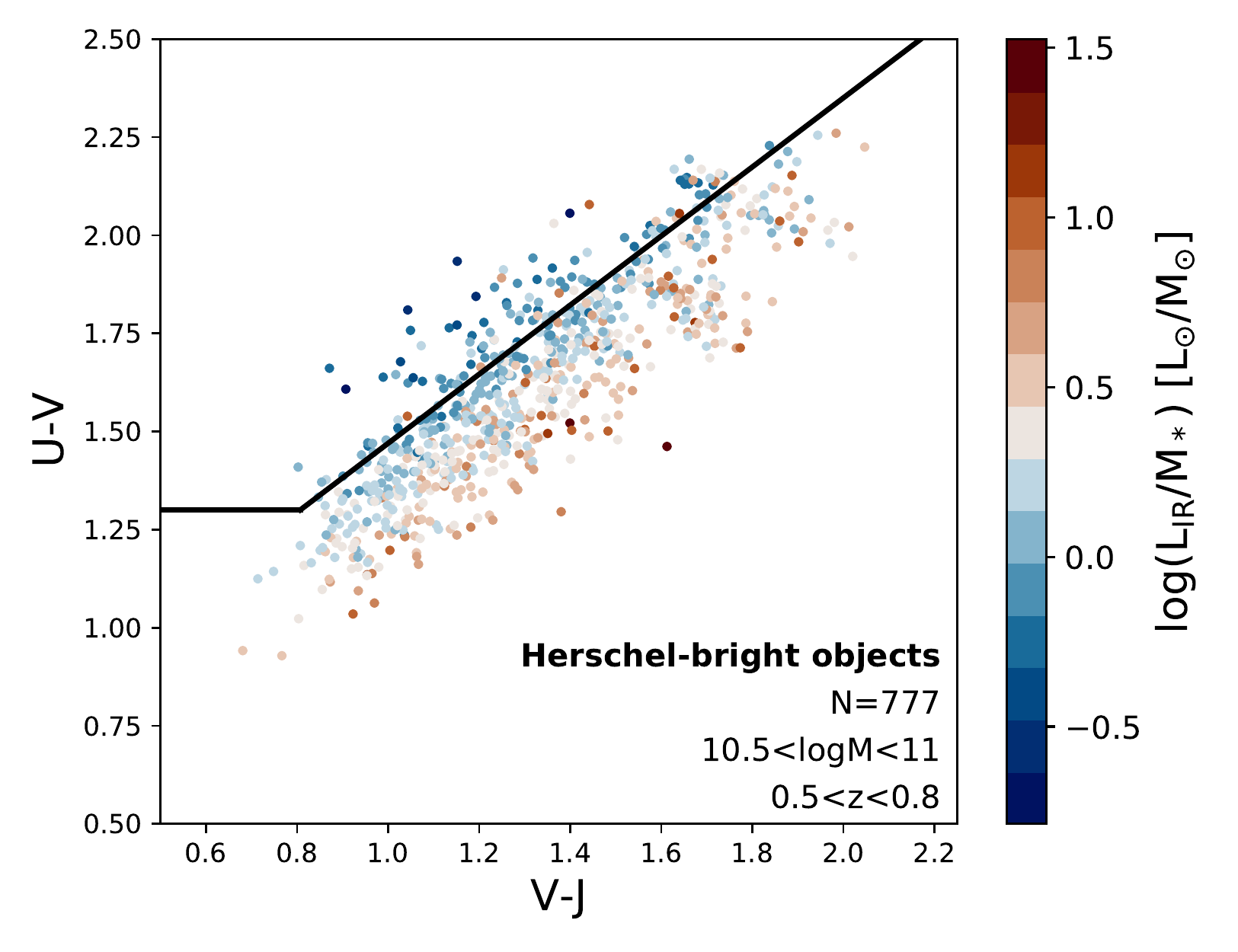}
\includegraphics[width=0.4\textwidth]{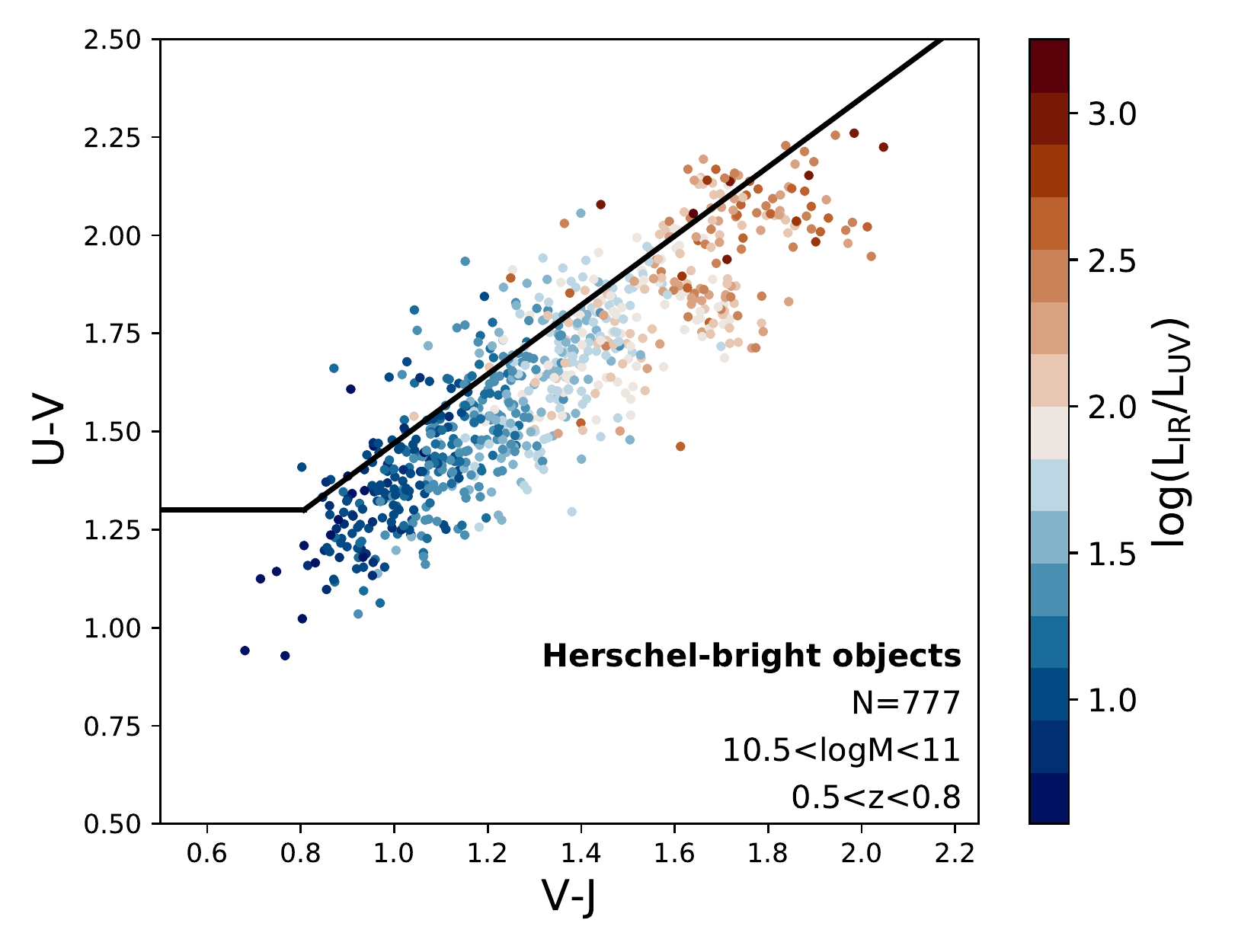}
\caption{Galaxies from the COSMOS2015 catalog in the $UVJ$ plane, color-coded by \LIR/M$_*$ (left) and \LIR/\LUV\ (right). Galaxy properties are derived by fits to the UV-to-IR photometry \citep{Leja_2020}. Only galaxies with at least two $Herschel$ detections are shown.}
\label{fig:uvjHerschel}
\end{figure*}

\subsection{Dust attenuation versus dust mass}

The main result of this work is that the observed variation in UVJ colors of star-forming systems at a given stellar mass can be attributed almost entirely to variation in viewing angle rather than variation in intrinsic dust mass. This brings several implications for galaxy studies. Most importantly, we conclude that the $UVJ$ diagram is not a good tool for identifying ``dusty'' galaxies. Colors that appear reddened might only indicate higher inclinations (or thicker galaxies, as illustrated in Figure~\ref{fig:geometricModel}), and not intrinsically higher dust content. On the other hand, if color is a useful indicator of inclination, our model could be used to select galaxies of a given inclination solely based on the photometry, instead of more difficult morphological measurements.

\subsection{The interpretation of IR and UV emission}

Another important consequence of our results is related to the use of \LIR/\LUV\ (i.e., the ratio of IR and UV luminosity) to estimate the proportion of star formation that is obscured. This method is based on the assumption that \LIR\ represents the obscured star formation while \LUV\ traces the unobscured star formation. 
However, in our model the dust attenuation in the UV is strongly dependent on inclination, while the IR luminosity is not affected by inclination since dust attenuation in the IR is negligible. This means that the ratio \LIR/\LUV\ at a given stellar mass is mostly a measurement of inclination, and not of the intrinsic fraction of obscured star formation. For example, adopting $D=0.25$, $\Avo=2.5$, and the Calzetti law, our model predicts a difference in the edge-on versus face-on attenuation of 4 magnitudes in the UV, meaning that the \LIR/\LUV\ ratio can vary by a factor of $\sim40$ solely due to inclination effects.

We test this prediction of our model using a small sample of galaxies with available far-IR data. We select galaxies from the COSMOS2015 catalog \citep{Laigle_2016} with $0.5 < z < 0.8$ and $10.5 < \log(M_\ast / M_\odot) < 11$ that are detected in \textit{Herschel}. Figure~\ref{fig:uvjHerschel} shows this sample on the $UVJ$ diagram, color-coded by  \LIR/M$_*$ (left) and \LIR/\LUV\ (right). Moving along the diagonal direction towards redder, more dust-attenuated systems, the \LIR/\LUV\ ratio varies by almost two orders of magnitude, while the total infrared luminosity (normalized by stellar mass) is virtually constant. This finding is consistent with our main result: at a given stellar mass, a galaxy's position along the dust vector in $UVJ$ space is determined primarily by galaxy orientation and does not correlate with the true amount of dust in the galaxy (see also \citealt{Arnouts_2013} for a similar analysis).

The fact that inclination has dramatically different effects on \LIR\ and \LUV\ should be taken into account in many other contexts; for example, when deriving star formation rates using the ``UV+IR'' method. Similarly, when imposing energy balance in photometric fits that include both UV and IR observations, if the inclination angle can be estimated from imaging then it is possible, in principle, to calculate inclination-dependent corrections (see \citealt{Doore_2021} for a first attempt in this direction). 

\subsection{Understanding the $UVJ$ colors of galaxies}

Advanced cosmological simulations are still unable to reproduce the full extent of the $UVJ$ colors of star-forming galaxies, particularly at the red end \citep{Donnari_2019, Akins_2021}. Our results can greatly help in understanding this discrepancy, and suggest that both the line-of-sight used to calculate the colors and the galaxy shapes are crucial for a meaningful comparison. Moreover, the fact that we are able to reproduce most of the observations using stochastic SFHs suggests that galaxy colors have limited constraining power on the history of star-forming galaxies, in agreement with the results of recent studies \citep{Chaves-Montero_2020, Hahn_2021}.

Our work can be considered the first step towards a fully quantitative understanding of the distribution of galaxies on the $UVJ$ diagram. Future developments should include a more realistic treatment of SFH, stellar metallicity, dust extinction law, and galaxy geometry, and should be extended to the quiescent population. Such a comprehensive empirical model of how galaxy colors evolve as a function of mass and redshift will provide new constraints on the formation, growth, and quenching of galaxies.

\section{Acknowledgements}
\label{sec:ack}

We acknowledge support from the SAO REU program, which is funded in part by the National Science Foundation REU and Department of Defense ASSURE programs under NSF  Grant no.\ AST-1852268,  and by the Smithsonian Institution. S.B. acknowledges support from the Clay Fellowship. S.T. is supported by the Smithsonian Astrophysical Observatory through the CfA Fellowship and by the 2021 Research Fund 1.210134.01 of UNIST (Ulsan National Institute of Science \& Technology).

\bibliography{biblio_film}

\end{document}